\documentstyle[bbox,epsf]{aipproc}

\def\be{\begin{equation}}
\def\ee{\end{equation}}
\def\bea{\begin{eqnarray}}
\def\eea{\end{eqnarray}}

\newcommand{\Gf}{GeV/fm$^3$ }
\newcommand{\Tc}{T_{\rm c}}
\newcommand{\vp}{\langle v_\perp\rangle}
\newcommand{\pT}{p_{\perp}}
\newcommand{\mt}{m_{\perp}}

\begin{document}

\title{From SPS to RHIC: Breaking the Barrier to the Quark-Gluon Plasma}

\author{Ulrich Heinz}
        
\address{Physics Department, The Ohio State University, Columbus, 
         OH 43210, USA\\
         Email: heinz@mps.ohio-state.edu}
       
\maketitle

\vspace*{-0.5cm}
\begin{abstract}
After 15 years of heavy-ion collision experiments at the AGS and SPS, 
the recent turn-on of RHIC has initiated a new stage of quark-gluon 
plasma studies. I review the evidence for deconfined quark-gluon matter 
at SPS energies and the recent confirmation of some of the key ideas 
by the new RHIC data. Measurements of the elliptic flow at RHIC provide 
strong evidence for efficient thermalization during the very early 
partonic collision stage, resulting in a well-developed quark-gluon 
plasma with almost ideal fluid-dynamical collective behaviour and a 
lifetime of several fm/$c$.  
\end{abstract}
\vspace*{-0.2cm}

\begin{center}
  {\bf I. \ THE QUARK-HADRON TRANSITION}
\end{center}
\label{sec1}

Quantum Chromodynamics (QCD), the theory of strong interactions, predicts 
for strongly interacting bulk matter a phase transition from a gas of 
hadron resonances (HG) at low energy densities to a quark-gluon plasma 
(QGP) at high energy densities. The critical energy density 
$\epsilon_{\rm c}$ is of the order of 1\,\Gf. It can be reached by 
heating matter at zero net baryon density to a temperature of 
about $\Tc\approx 170$\,MeV, or by compressing cold nuclear matter
to baryon densities of about $\rho_{\rm c}\sim 3-10\,\rho_0$ (where 
$\rho_0=0.15$\,fm$^{-3}$ is the equilibrium density), or by combinations 
thereof. A simple version of the phase diagram is shown in Fig.~1 
\cite{Raja}.
 
By colliding heavy ions at high energies one hopes to create hadronic 
matter at energy densities above $\epsilon_{\rm c}$. At lower energies 
(SIS{\,@\,}1\,$A$\,GeV), the nuclei are stopped, compressed and 
moderately heated. At higher energies (AGS{\,@\,}10\,$A$\,GeV and 
SPS @ 160\,$A$\,GeV) one reaches higher temperatures, but the 
colliding nuclei are no longer completely stopped and the baryon 
chemical potential of the matter created at rest in the c.m.s. decreases. 
At the colliders RHIC ($\sqrt{s}=200\,A$ GeV) and LHC 
($\sqrt{s}=5500\,A$\,GeV) the baryon chemical potential is so small that 
one essentially simulates the nearly baryon-free hot hadronic matter of
the early universe. If the matter thermalizes quickly at energy densities 
above $\epsilon_{\rm c}$, it will pass through the {\em quark-hadron phase 
transition} as the collision fireball expands and cools.

Along the temperature axis at $\mu_{\rm B}=0$ our knowledge of the 
QCD phase diagram is based on hard theory (lattice QCD) \cite{Karsch}, 
but for nonzero baryon density we must still rely on models interpolating 
between low-density hadronic matter, described by low-energy effective 
theories, and high-density quark-gluon plasma, described by perturbative 
QCD \cite{Raja}. The uncertainties at high-baryon densities are thus 
relatively large (typically ${\cal O}(30-50\%)$). At zero baryon density, 
numerical simulations of QCD with 3 dynamical light quark flavors on the 
lattice are now available, and the systematic errors due to lattice 
discretization and continuum extrapolation are beginning to get small 
\cite{Karsch}. The critical temperature $\Tc$ for real-life QCD is 
predicted as $T_{\rm c} \approx 170$\,MeV\,$\pm10\%$ \cite{Karsch}. Near 
$T_{\rm c}$ the energy density in units of $T^4$ changes dramatically by 
more than a factor of 10 within a very narrow temperature interval. Above 
$T\simeq 1.2\, \Tc$, $\epsilon/T^4$ appears to settle at about 80\% of 
the Stefan-Boltzmann value for an ideal gas of non-interacting quarks 
and gluons. 

\begin{minipage}[c]{8cm}
\vspace*{0.3cm}
\hspace*{-0.5cm}
  \epsfxsize 8cm \epsfysize 7.7cm
  \epsfbox{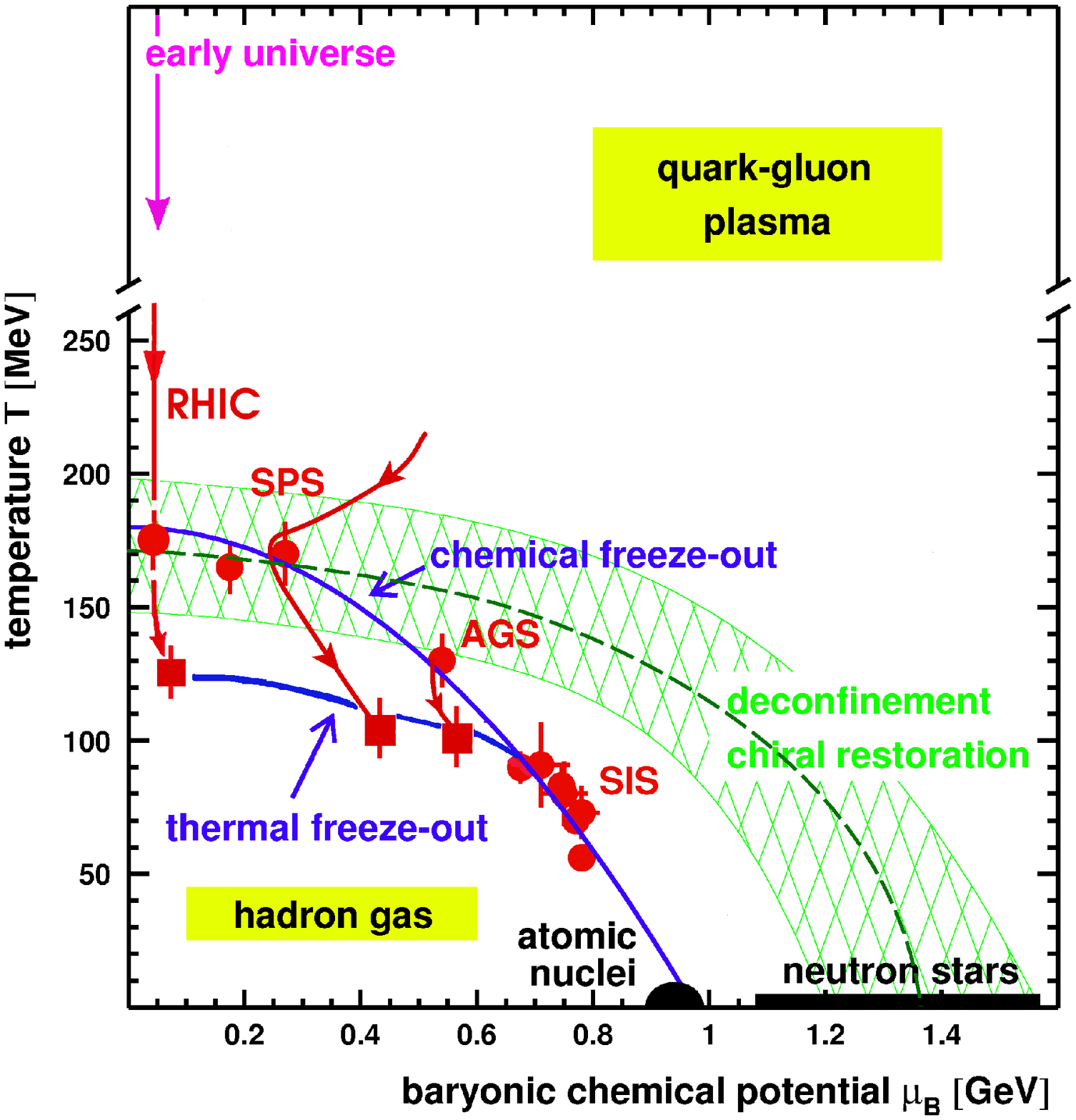}
\end{minipage}
\hfill\hspace*{-1.5cm}
\parbox[c]{6.5cm}{\vspace*{-0.1cm}
    {\small FIGURE~1. Sketch of the QCD phase diagram, temperature $T$ vs. 
    the chemical potential $\mu_{\rm B}$ associated with net baryon 
    density $\rho_{\rm B}$. The cross-hatched region indicates the 
    expected phase transition and its present theoretical uncertainty,
    the dashed line representing its most likely location. Lines
    with arrows indicate expansion trajectories of thermalized matter 
    created in different environments. 
    For a discussion of the chemical and thermal freeze-out lines and 
    the location of the data points see text.}
\label{F1}}
\vspace*{0.2cm}

According to lattice QCD, only about 0.6\,GeV/fm$^3$ of energy 
density are needed to make the transition to deconfined quark-gluon 
matter \cite{Karsch}. But it is very expensive to reach temperatures 
well above $\Tc$: an initial temperature of 220 MeV, ${\approx}30\%$ 
above $\Tc$, already requires an initial energy density 
$\epsilon\simeq 3.5$\,\Gf$\!$, about 6 times the critical value. This 
limits severely the reach of the CERN SPS into the QGP phase: only the 
region at and slightly above $\Tc$ can be probed. But with RHIC 
considerably higher energy densities have now become accessible so that 
we can penetrate deeper into the new phase. The situation will further 
improve with the beginning of the heavy-ion program at the LHC at CERN 
in 2007.

At $\Tc$ two phenomena occur simultaneously \cite{KL94}: color
confinement is broken, i.e. colored degrees of freedom can propagate 
over distances much larger than the size of a hadron, and the 
approximate chiral symmetry of QCD, which is spontaneously broken at
low temperatures and densities, gets restored. Both effects 
significantly accelerate particle production: the liberation of gluons 
in large densities opens up new gluonic production channels, and the 
threshold for quark-antiquark pair production is lowered due to the
restoration of the spontaneously broken chiral symmetry at low densities. 

\begin{center}
{\bf II. \ RECONSTRUCTING THE LITTLE BANG}
\end{center}

As the two nuclei hit each other, a superposition of nucleon-nucleon 
(NN) collisions occurs. What is different from individual NN collisions 
is that (i) each nucleon scatters several times, and (ii) the partons 
liberated in different NN collisions rescatter with each other even 
before hadronization, as do the hadrons afterwards. Both change the 
particle production {\em per participating nucleon}, but only the 
rescattering processes can lead to a state of local thermal equilibrium, 
by redistributing the energy lost by the beams into the statistically 
most probable configuration. They cause thermodynamic pressure acting 
against the outside vacuum, and this makes the reaction zone expand 
collectively. {\em This is a genuine collective nuclear effect, with 
no analogue in elementary particle collisions}. The expansion dilutes 
the fireball below $\epsilon_{\rm c}$, at which point hadrons are formed 
from the quarks and gluons (hadronization). Further interactions among 
these hadrons cease once their average distance exceeds the range of the 
strong interactions: the hadrons ``freeze out''.

The strong interactions among the partons and hadrons before freeze-out 
wipe out much information about their original production processes.
Extracting information about the hot and dense early collision stage 
thus requires to exploit features which are either established early 
and survive the rescattering and collective expansion or can be 
reliably back-extrapolated. Correspondingly one classifies the 
observables into two classes, {\em early} and {\em late} signatures. 

The conceptually cleanest early signatures are the directly produced 
real and virtual photons since these re-interact weakly and escape 
directly from the fireball (virtual photons materialize as $e^+e^-$ or 
$\mu^+\mu^-$ pairs). They are emitted throughout the expansion, but
their production should be strongly weighted towards the hot and dense 
initial stages. Unfortunately, direct photons are rare, and the 
experimental background from hadronic decay photons after freeze-out 
is enormous. 

Another early signature are hadrons containing charmed quarks. At the SPS, 
$c\bar c$ pairs can be only created in the primary NN collisions; secondary 
scatterings are already below the $c\bar c$ threshold. The latter can only 
redistribute them in phase-space, changing the relative amounts of mesons 
with hidden and open charm. It was shown that such a redistribution is much 
easier in the color-deconfined QGP phase than by reinteractions of charmed 
and other hadrons; in this way charm-redistribution becomes also an early 
signature. At SPS energies and below, charm production is a 
very rare process and only hidden charm mesons ($J/\psi,\psi'$) have 
been measured. The interpretation of what happens to them crucially
depends on the assumption that no secondary charm is produced. At RHIC 
energies and above the production of additional charmed particles in 
secondary collisions becomes possible, and a consistent interpretation
of charmonium data requires also the measurement of open charm. None of 
the present RHIC experiments can do that.

Hadrons made of $u$ and $d$ quarks can be easily produced and destroyed 
throughout the fireball expansion. Their abundances and spectra are 
{\em late signatures} which provide only {\em indirect} information 
about the early collision stages. But they are very numerous and can 
be measured very accurately. We'll use these {\em late} signals to 
reconstruct the Little Bang and then check the consistency of the 
resulting picture with the less detailed direct information from the 
{\em early} signals. Hadrons involving strange quarks play an intermediate 
role: $s\bar s$ pairs are easily produced in the dense color-deconfined, 
chirally almost symmetric QGP phase, but hadron interactions after 
hadronization leave their abundances essentially unaltered. Their yields
thus reflect the situation reached at the quark-hadron transition point. 

It was recently realized that in {\em non-central} collisions the observed 
collective flow pattern of {\em all} hadrons shows certain anisotropies 
(``elliptic flow'') which are established very early in the collision, 
even before hadrons are formed, and are hardly changed in the late expansion 
stages. The elliptic flow of the emitted hadrons (including the abundant 
light ones) thus constitutes another {\em early collision signature}. 

\vspace*{0.2cm}
\begin{center}
{\bf III. \ INITIAL CONDITIONS}
\end{center}

We can estimate the initially {\em produced} energy density by dividing
the measured total transverse energy $E_T$ by an estimated initial 
reaction volume \cite{Bj} 
 \begin{equation}
 \label{bj}
  \epsilon_{\rm Bj}(\tau_0) =
  {1\over \pi R^2}\, {1\over \tau_0} \, {dE_T\over dy}\, .
 \end{equation}
Here $\tau_0\,dy$ is an estimate for the length of a cylinder undergoing
boost-invariant longitudinal expansion \cite{Bj} from which particles
with longitudinal momenta in a rapidity interval $dy$ are emitted.  
Inserting for $\pi R^2$ the overlap area of two Pb nuclei 
colliding at zero impact parameter, choosing $\tau_0=1$\,fm/$c$, and
using $dE_T/dy(y=0)\approx 400$\,GeV for central Pb+Pb collisions 
\cite{epsilon0} gives
 \begin{equation}
 \label{eps}
  \epsilon_{\rm Bj}^{\rm Pb+Pb}(1\,{\rm fm}/c) =
  3.2 \pm 0.3\, {\rm GeV/fm}^3\,.
 \end{equation}
Note that this is the average over the transverse plane; the value 
$\epsilon_0{\,=\,}\epsilon(r{=}0)$ in the center is about twice as high. 
The analogous value extracted from $\sqrt{s}{\,=\,}130\,A$\,GeV Au+Au 
collisions at RHIC \cite{PHENIX} is 60{\%} larger, and recent PHOBOS 
data from $\sqrt{s}{\,=\,}200\,A$\,GeV show another increase by about 15{\%}
\cite{PHOBOS}. As QGP searchers we are clearly playing in the right 
ball-park: if the matter were already thermalized after 1\,fm/$c$ (for 
which there is evidence from elliptic flow measurements, see below), the 
temperature corresponding to the SPS value (\ref{eps}) would be 
$T_0\simeq 210-220$\,MeV. At RHIC, thermalization seems to happen even 
earlier, at around 0.5-0.6\,fm/$c$ \cite{KH3}, and the initial temperature 
may have been as high as 350 MeV.

\vspace*{0.2cm}
\begin{center}
{\bf IV. \ THERMAL FREEZE-OUT: AN EXPLODING FIREBALL}
\end{center}

The measured hadron spectra contain two pieces of information: (i) Their
normalizations, i.e. the {\em yields and abundance ratios}, provide 
the chemical composition of the fireball at the ``chemical freeze-out'' 
point where the hadron abundances freeze out; this gives information 
about the degree of chemical equilibration, see Sec. VI. (ii) The 
hadronic {\em momentum spectra} provide information about thermalization 
of the momentum distributions and collective flow. The latter is caused 
by thermodynamic pressure and thus reflects, in a time-integrated way, 
the equation of state of the fireball matter. We concentrate on the 
transverse flow since all of it is generated {\em during} the reaction.  

In a thermal, collectively expanding system the shapes of {\em all} 
hadronic $\mt$-spectra ($\mt{\,=\,}\sqrt{m^2{+}\pT^2}$) can be 
characterized by just two numbers: the temperature $T_{\rm f}$ and 
the mean transverse flow velocity $\vp$ at freeze-out. This is true 
if all hadrons decouple simultaneously, i.e. if their rescattering cross 
sections are similar. This can be checked experimentally: e.g., it was 
observed that the $\Omega$ \cite{WA97spec} and $J/\psi$ 
\cite{NA50slope} show steeper slopes than expected from the systematics
of the remaining hadrons, because they have smaller hadronic rescattering
cross sections and thus decouple earlier \cite{Hecke}. For all other 
hadrons, a common parametrization by a single pair ($T_{\rm f},\vp$)
works very well. The transverse flow velocity $\vp$ manifests
itself as a flattening of the $\mt$-spectra, by a mass-independent 
blueshift factor $T_{\rm slope}{\,=\,}T_{\rm f}\sqrt{(1{+}\vp)/(1{-}\vp)}$
in the relativistic domain $\mt{\,>\,}2m_0$ and by a mass-dependent
term $T_{\rm slope}{\,=\,}T_{\rm f}+{\textstyle{1\over 2}}m_0\vp^2$
at small $\pT$. A roughly linear mass-dependence of the transverse 
slopes at low $\pT$ was observed at the SPS (see e.g. \cite{WA97spec}).
Fig.~2 shows that the RHIC data \cite{PHENIX_spec,STAR_spec} follow the 
same systematics, with antiproton spectra being much flatter at low 
$\pT$ than the pion spectra, in quantitative agreement with hydrodynamic 
calculations which assume complete thermalization of the fireball after 
$\tau_0=0.6$\,fm/$c$ \cite{HKHRV}. 

\begin{minipage}[c]{8cm}
\vspace*{0.5cm}
\hspace*{-0.5cm}
  \epsfxsize 7.8cm \epsfysize 7cm
  \epsfbox{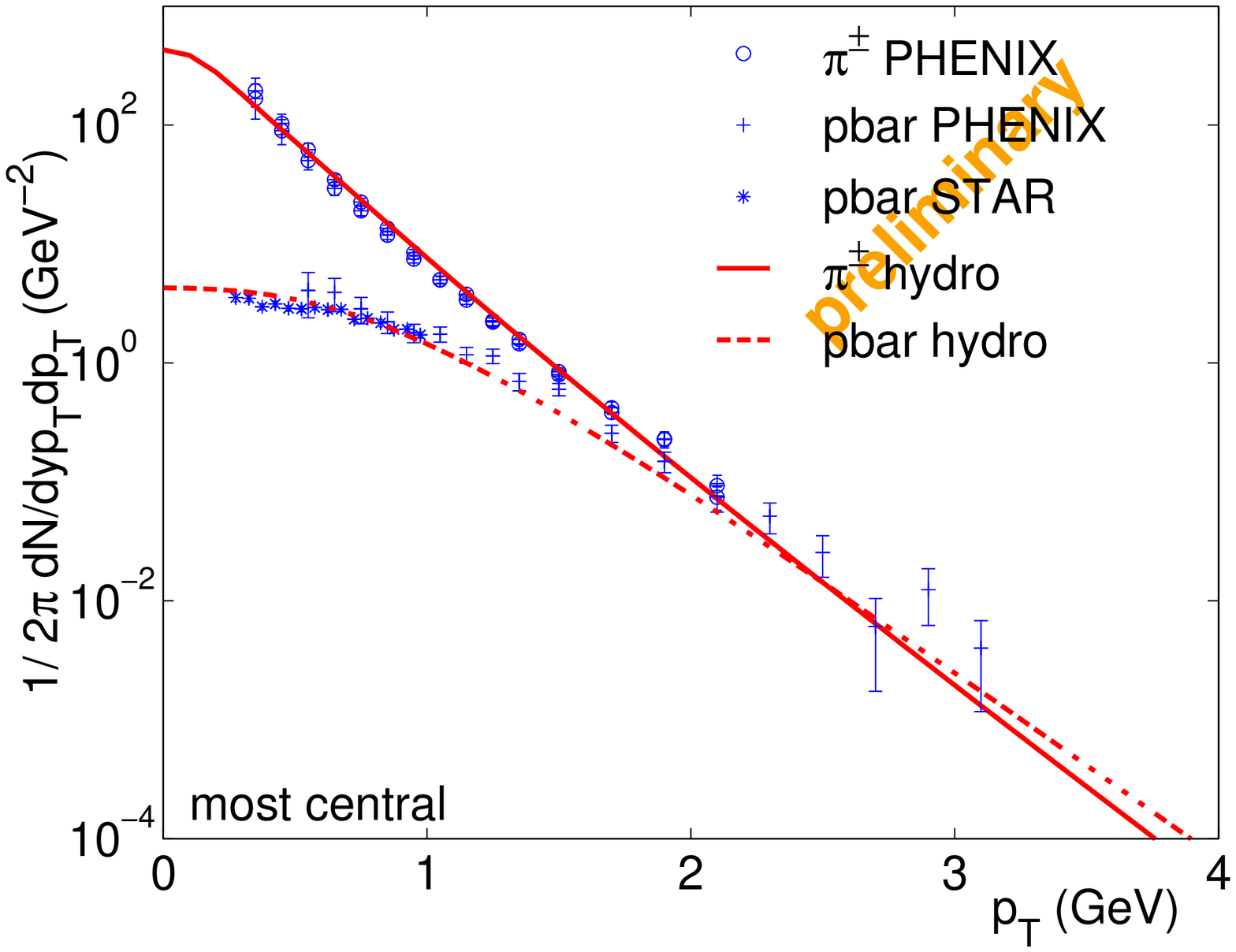}
\end{minipage}
\hfill\hspace*{-1.5cm}
\parbox[c]{6.5cm}{\vspace*{-0.3cm}
    {\small FIGURE~2. Preliminary transverse momentum spectra of charged 
    pions and antiprotons from 130\,$A$\,GeV Au+Au collisions at RHIC 
    \protect\cite{PHENIX_spec,STAR_spec}, compared with hydrodynamic 
    predictions \protect\cite{HKHRV}, corresponding to 
    $T_{\rm f}=128$\,MeV, $\mu_B=70$\,MeV, and $\vp=0.6\,c$ (private 
    communication by P.F. Kolb whom I thank for preparing this figure).}
\label{F2}}
\vspace*{0.2cm}

Transverse flow also affects two-particle momentum correlations
and gives, for example, rise to a characteristic $\mt$-dependence of
the source sizes extracted from Bose-Einstein correlation measurements
\cite{HBT} or deuteron yields \cite{coal,NA44,NA52}. The most accurate 
separation of thermal and collective contributions to the final hadron 
momenta is obtained from a simultaneous analysis of single-particle 
spectra and two-particle correlations \cite{NA49HBT,TWH99}. At the SPS 
this gives $T_{\rm f}\approx 100$\,MeV and $\vp\approx 0.55\,c$. The 
corresponding freeze-out energy density is only about 50\,MeV/fm$^3$!
For RHIC such a combined analysis is not yet available, but the spectral 
slopes alone (see Fig.~2) indicate a somewhat higher freeze-out temperature 
$T_{\rm f}\approx 125$\,MeV, with a 20{\%} larger radial flow at RHIC 
than at the SPS at the same value of $T$.

\vspace*{0.2cm}
\begin{center}
{\bf V. \ THE MISSING RHO: WATCHING THERMALIZATION}
\end{center}

If the Little Bang started at initial energy densities above 3\,\Gf, but 
decoupled only at about 50\,MeV/fm$^3$, how can we find out what happened 
in between? The $\rho$ meson provides a first answer: it can decay into 
$e^+e^-$ or $\mu^+\mu^-$ pairs which escape from the fireball without 
further interactions, and this $\rho$-decay clock ticks at a rate of 
1.3\,fm/$c$, the $\rho$ lifetime. What I mean by this is that after one 
generation of $\rho$'s has decayed, a second generation is created by 
resonant $\pi\pi$ scattering, which can again decay into dileptons, etc. 
The number of extra dileptons with the invariant mass of the $\rho$ is 
thus a measure for the time in which the fireball consists of strongly 
interacting hadrons \cite{HL91}. Obviously, $\rho$ mesons do not exist 
before hadrons appear in the fireball, so they won't tell us anything 
about a possible initial QGP phase. But they still allow us to look 
{\em inside} the strongly interacting hadronic fireball at a later stage, 
still long before the hadrons decouple.

\begin{minipage}[c]{8cm}
\vspace*{-0.5cm}
\hspace*{-0.6cm}
  \epsfxsize 8cm \epsfysize 8.5cm 
  \epsfbox{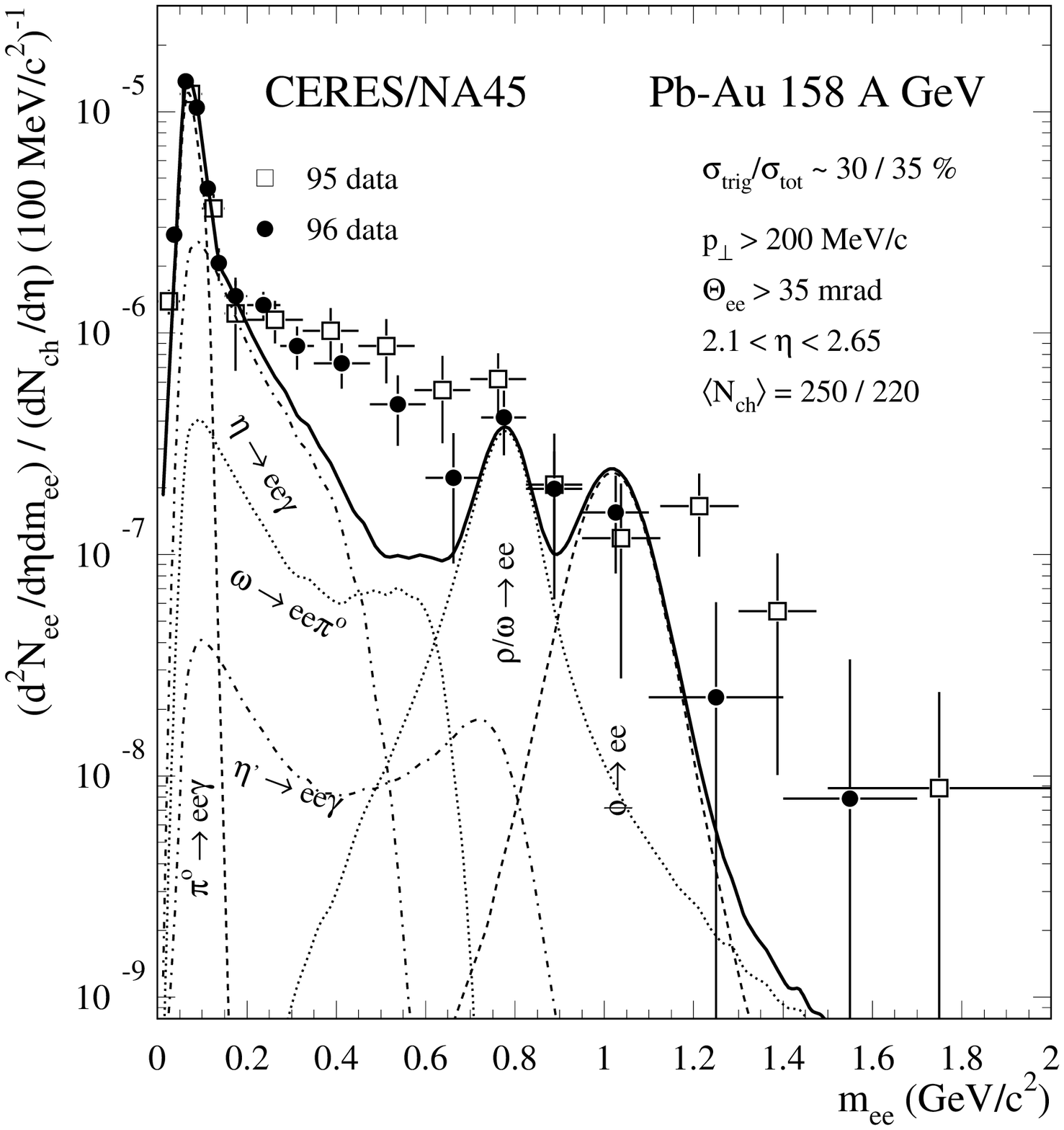}
\end{minipage}
\hfill\hspace*{-2cm}
\parbox[c]{6.5cm}{\vspace*{0.3cm}
    {\small Figure~3. Invariant mass spectrum of $e^+e^-$ pairs from 
    158\,$A$\,GeV/$c$ Pb+Au collisions \protect\cite{CERES}. The solid line
    is the expected spectrum (the sum of the many shown contributions)
    from the decays of hadrons produced in $pp$ and $pA$ collisions 
    (where it was experimentally checked \protect\cite{CERES}), properly 
    scaled to the Pb+Au case. Two sets of data with different analyses 
    are shown. Note that the $\rho$-peak reappears if only $e^+e^-$ 
    pairs with $p_\perp>500$\,MeV/$c$ are selected \protect\cite{CERES}; such 
    fast $\rho$'s escape quickly from the fireball and are not as 
    strongly affected by collision broadening.}
\label{F3}}
\vspace*{0.15cm}

However, when the CERES/NA45 collaboration looked at the $e^+e^-$ spectrum 
in 158\,$A$\,GeV/$c$ Pb+Au collisions (see Fig.~3) and couldn't find the 
$\rho$ at all! They saw extra $e^+e^-$ pairs in the mass region of the 
$\rho$ and below (about 2.5--3 times as many as expected), but instead of 
a nice $\rho$-peak at $m_\rho=770$\,MeV one sees only a broad smear 
\cite{CERES}. Many explanations of the CERES-effect have been proposed, 
but the simplest one consistent with the data (for a review see \cite{Rapp}) 
is {\em collision broadening}: there is strong rescattering of the pions, 
not only among each other, but also with the baryons in the hadronic 
resonance gas, and this modifies their spectral densities and, 
as a consequence, leads to a smearing of the $\rho$-resonance in 
the $\pi\pi$ scattering cross section. 

This demonstrates that, after first being formed in the hadronization 
process, the pions (the most abundant species at the SPS) undergo 
intense rescattering before finally freezing out. And this 
again is the mechanism which allows the fireball to reach and maintain 
a state of approximate local thermal equilibrium, to build up 
thermodynamic pressure and to collectively explode, as seen from 
the above analysis of the freeze-out stage. That the dileptons from
collision-broadened $\rho$'s outnumber those from the decay of 
unmodified $\rho$'s emitted at thermal freeze-out (which should show 
up as a normal $\rho$-peak) shows that the hadronic rescattering 
stage must have lasted several $\rho$ lifetimes. 

\vspace*{0.2cm}
\begin{center}
{\bf VI. \ SEEING THE QUARK-HADRON TRANSITION}
\end{center}

In the rest of this talk I will concentrate on observables which were 
found to differ drastically in AA and NN collisions but which we now 
believe cannot be changed efficiently by hadronic rescattering during 
the time between hadronization and kinetic freeze-out. Observables for 
which this can be firmly established yield insights about where AA and 
NN collisions differ already {\em before or during} hadronization, 
irrespective whether or not the hadrons rescatter after being formed. 

One possibility how the early stage of an AA collision may differ from 
that in NN collisions is the formation of a quark-gluon plasma. I 
therefore review a few key QGP predictions and check how they fare in 
comparison with the data. In the present Section I discuss {\em 
strangeness enhancement} as a QGP signature, returning to two further 
QGP predictions in the following two Sections. 

{\em Strangeness enhancement and chemical equilibration} were among
the earliest predicted QGP signatures \cite{Raf}. The idea is simple:
1. Color deconfinement leads to high gluon density, fostering $s\bar s$ 
creation by gluon fusion. 2. Chiral symmetry restoration renders the 
$s$-quarks relatively light, and in the QGP they can be crea\-ted without 
the need for additional light quarks to make a hadron; both effects lower
the production threshold. Both should considerably reduce the time needed 
for strangeness saturation and chemical equilibration compared to hadronic 
rescattering processes. Since in NN and $e^+e^-$ collisions strange hadron 
production is known to be suppressed relative to simple phase-space 
considerations \cite{Becattini}, this should cause a relative 
{\rm strangeness enhancement} in heavy-ion collisions.

Kinetic simulations based on known hadronic properties and interaction 
cross sections have shown that it is impossible to create a state of 
hadronic chemical equilibrium and a significant amount of strangeness 
enhancement out of a non-equilibrium initial state by purely hadronic 
rescattering \cite{Hqm99}. If you want to get these features out, you 
have to put them in at the beginning of the simulation. 

There may be many different ways of doing so, but the most efficient way 
of creating a state of (relative or absolute) hadronic chemical equilibrium 
may be provided by the hadronization process itself: If before 
hadronization the quarks and gluons are uncorrelated (such as in a 
QGP), then the most likely outcome of the non-perturbative hadronization 
process is a statistical occupation of the hadronic phase-space, a state 
of maximum entropy. If (as predicted for the QGP \cite{Raf}) the number of 
$s\bar s$-pairs is enhanced {\em before} the onset of hadronization, or 
by the fragmentation of gluons {\em during} hadronization, their 
statistical distribution over the available hadronic channels will 
naturally lead to an apparent hadronic chemical equilibrium state 
(with the corresponding enhancement of, say, the $\bar \Omega$) 
{\em even if none of the hadrons ever scattered with each other after 
being formed}.

\begin{minipage}[c]{8cm}
\vspace*{-0.5cm}
\hspace*{-0.6cm}
  \epsfxsize 8cm \epsfysize 7cm
  \epsfbox{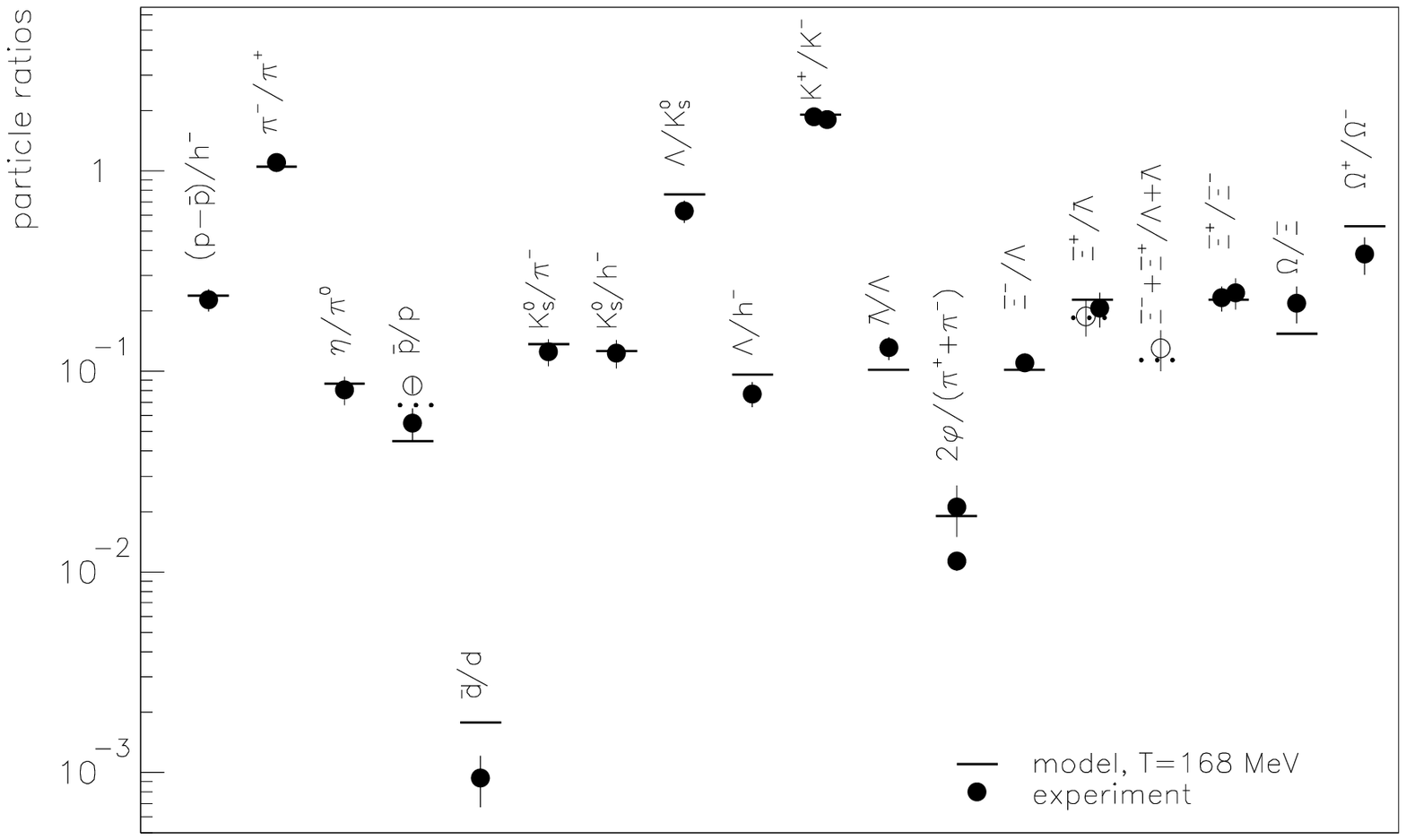}
\end{minipage}
\hfill\hspace*{-1.5cm}
\parbox[c]{6.5cm}{\vspace*{-0.5cm}
    {\small FIGURE~4. A compilation of measured particle ratios from 
    158\,$A$ GeV/$c$ Pb+Pb collisions, compared with a hadron resonance
    gas in complete chemical equilibrium (full strangeness saturation)
    at $T_{\rm chem}=168$\,MeV and $\mu_{\rm B}=266$\,MeV \protect\cite{PBM}.}
\label{F4}}
\vspace*{-0.4cm}

Such a state of ``apparent'' or ``pre-established'' chemical equilibrium 
is indeed seen in the experiments: Fig.\,4 shows 18 hadronic particle 
ratios from 158\,$A$ GeV Pb+Pb collisions, compared with a hadronic 
chemical equilibrium state at $T_{\rm chem}{\,=\,}168$\,MeV and 
$\mu_{\rm B}{\,=\,}266$\,MeV \cite{PBM}. A similar analysis of recent 
RHIC data \cite{PBMnew} gives $T_{\rm chem}\simeq 175$\,MeV
and $\mu_{\rm B}\simeq 50$\,MeV. The ratio of strange to non-strange
quarks in the final hadrons is about a factor 2 higher than in $pp$ 
collisions \cite{BGS}, and triply strange $\Omega$ baryons are enhanced 
relative to pBe collisions by a factor 17 \cite{Lietava}! This and the 
value of $T_{\rm chem}$ are interesting: $T_{\rm chem}$ characterizes the 
energy density at which hadronization occurs (about 0.5\,GeV/fm$^3$) 
and coincides with the critical temperature for color deconfinement 
from lattice QCD. If the hadrons were formed by hadronization at the 
critical energy density $\epsilon_{\rm c}$ and their abundances froze 
out at $T_{\rm chem}\simeq 170$\,MeV, there was no time to achieve this 
equilibrium configuration by hadronic rescattering; the hadrons must 
have been ``born'' into chemical equilibrium \cite{Hqm97,Stock}. Since 
subsequent hadronic rescattering is inefficient in changing the hadronic 
abundances \cite{Hqm99}, we can thus use them to measure $T_{\rm c}$.  

\vspace*{0.2cm}
\begin{center}
{\bf VII. \ $J/\psi$ SUPPRESSION AND COLOR DECONFINEMENT}
\end{center}

What is the nature of the state from which the hadrons emerge in this 
manner? This question brings us to the second key QGP prediction 
\cite{MS86}: the high gluon density resulting from color deconfinement 
should Debye-screen the color interaction between a $c$ and a $\bar c$ 
quark produced during the initial nuclear impact, thus preventing them 
from binding into charmonium states ($J/\psi$, $\chi_c$, $\psi$'). 
Instead, they would eventually find light quark partners to make hadrons
with open charm. The result should be a suppression of charmonium 
production in heavy-ion collisions, and the screening mechanism should 
lead to a specific suppression pattern which, as a function of the
achieved energy density, first affects the loosely bound $\psi$' and 
$\chi_c$ states and then the strongly bound $J/\psi$ ground state 
\cite{satz}.

Nuclear $J/\psi$ suppression was indeed found at the SPS by the NA38/NA50 
Collaboration. Fig.~5 shows that, as a function of collision centrality, 
measured by the produced transverse energy $E_T$ and in the left panel
translated into an energy density at $\tau=1$\,fm/$c$ using a generalization
of Eq.~(\ref{bj}), the yield of $J/\psi$ mesons (identified by their 
$\mu^+\mu^-$ decay) is suppressed ``anomalously'' (i.e. below 
expectations). For a discussion of ``normal'' $J/\psi$ suppression 
I must refer to Refs.~\cite{satz,NA50eps}; it is well-defined and
carefully experimentally tested and represented by the horizontal line 
in the left panel of Fig.~5. The observed deviation from this normal 
suppression is in qualitative agreement with the QGP prediction; 
in particular, the weakly bound $\psi'$ (not shown) already suffers 
anomalous suppression in central S+U collisions while the strongly bound 
$J/\psi$ shows it only in semicentral and central PbPb collisions. The 
observed pattern is not yet understood in all details \cite{satz,Bl}; 
however, it definitely cannot be reproduced by final state rescattering 
of the $J/\psi$ with the dense hadronic environment after hadronization, 
represented by four independent calculations shown as lines in the right 
panel of Fig.~5.

\vspace*{-0.3cm}
\hspace*{-1.2cm}
\begin{center}
   \begin{minipage}[t]{7truecm}
         \epsfxsize 7truecm \epsfysize 6truecm \epsfbox{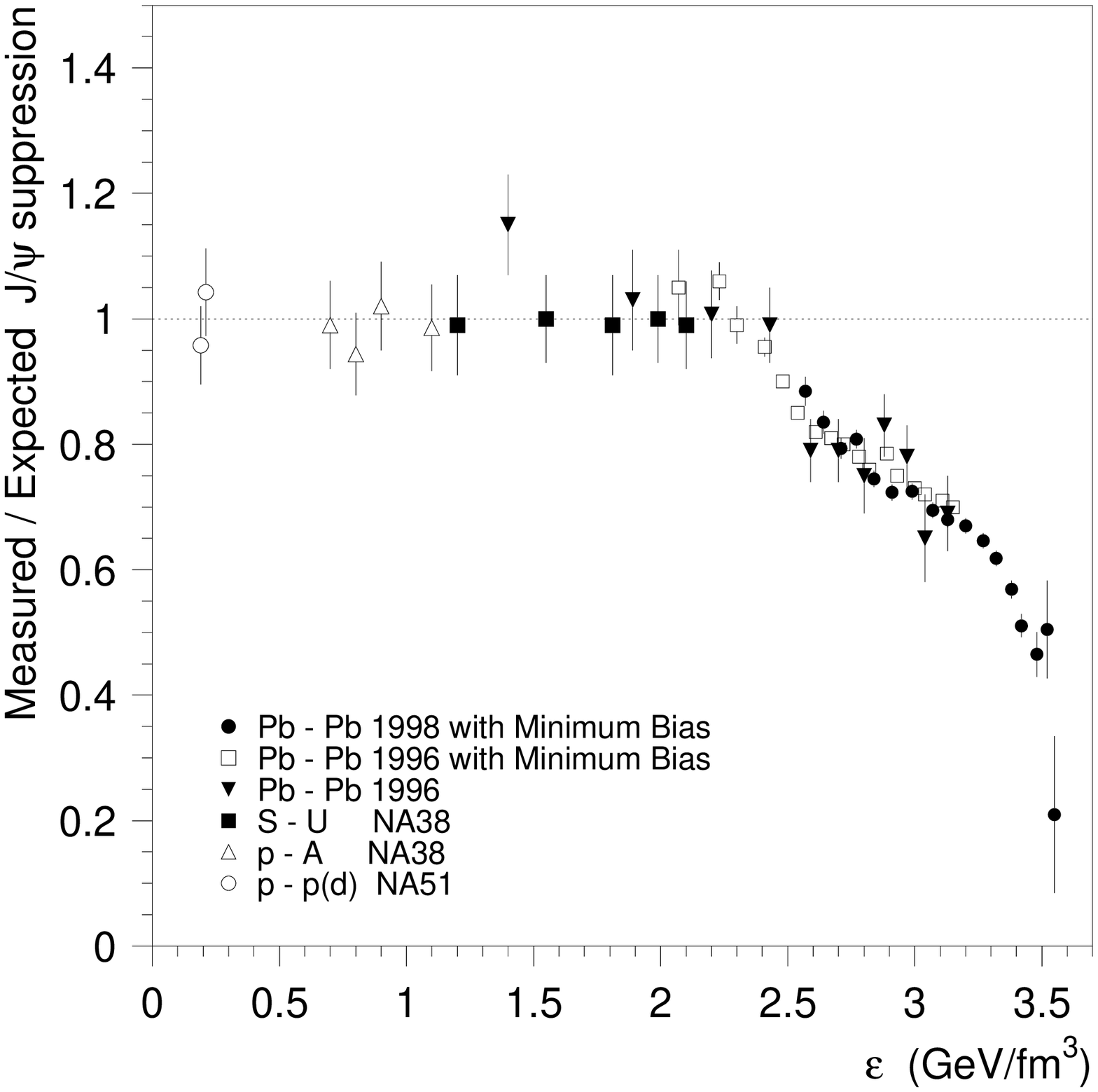}
         \hfill
   \end{minipage}
   \begin{minipage}[t]{7truecm}
         \epsfxsize 7truecm \epsfysize 6truecm \epsfbox{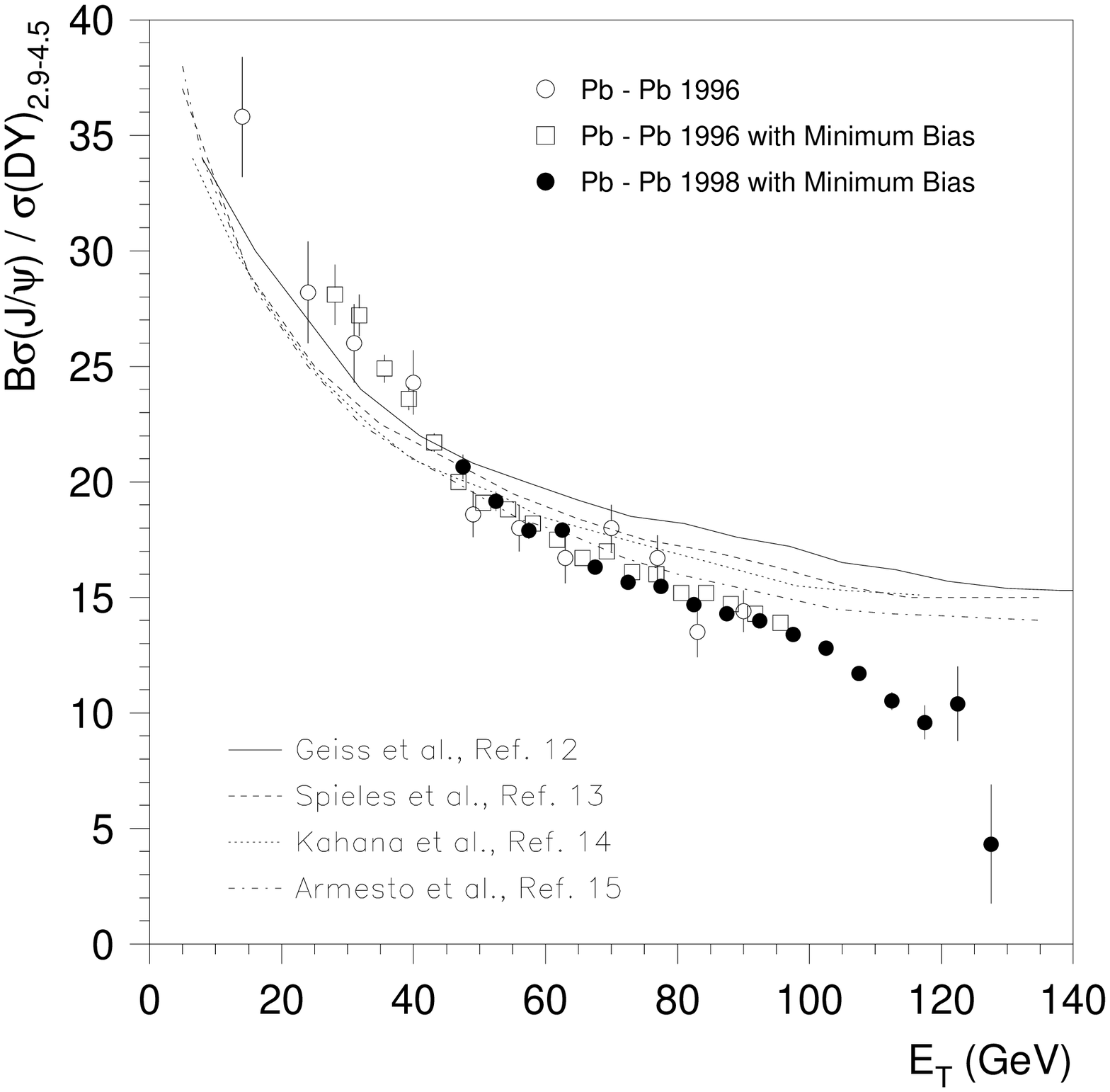}
         \hfill
   \end{minipage}
\end{center}
\vspace*{-0.7cm}
  {\small FIGURE~5. Left: ``Anomalous $J/\psi$ suppression'' as a function 
  of initial energy density \protect\cite{NA50eps}. Right: The ratio 
  $(J/\psi)$/Drell-Yan production, as a function of the measured 
  transverse energy $E_T$, compared to hadronic comover 
  models \protect\cite{NA50eps}.}
\label{F5}
\vspace*{0.2cm}

\begin{center}
{\bf VIII. \ THERMAL ELECTROMAGNETIC RADIATION}
\end{center}

The third (and earliest) key prediction for QGP formation is thermal
radiation of (real and virtual) photons from the thermalized quarks in 
the QGP \cite{Sh}. In spite of the difficulties arising from large 
backgrounds, experiments at the SPS have searched for such radiation. In
both the real \cite{WA98} and virtual photon ($\mu^+\mu^-$ \cite{NA50exc}) 
channels enhancements over hadronic decay backgrounds were reported.
However, no unambiguous connection with thermal radiation has been 
established, in part because the predicted signal for the latter
\cite{Gall} is marginal at the SPS, given the accuracy of the experiments. 
To see the plasma ``shine'' one needs the higher temperatures and longer 
plasma lifetimes which can be reached at RHIC.

\newpage

\begin{center}
{\bf IX. \ ELLIPTIC FLOW: AN EARLY HADRONIC SIGNATURE}
\end{center}

In non-central heavy-ion collisions the initial overlap region is 
spatially deformed into an almond shape in the transverse plane. 
Nevertheless, at each point $\bbox{r}$ the initial transverse momentum 
distribution is isotropic. If the produced matter expands without further 
interaction (``free streaming''), the $\pT$ distribution remains 
isotropic while the spatial deformation eventually disappears (the 
almond looks more spherical as it grows). On the other hand, if the 
produced matter thermalizes quickly, pressure builds up inside, and 
the spatial deformation results in anisotropic pressure gradients.
These generate stronger collective flow in the shorter direction (i.e. 
into the reaction plane) than in the longer one (i.e. out of the reaction 
plane) \cite{Olli}, and the $\pT$-distribution becomes anisotropic. This 
phenomenon is called elliptic flow, measured by the second coefficient 
$v_2$ of an azimuthal Fourier decomposition of the $\pT$-spectrum 
\cite{VZ96}. Elliptic flow lets the almond grow faster along its short 
than along its long direction, thereby reducing its deformation. When 
the spatial deformation and the accompanying anisotropic pressure 
gradients vanish, $v_2$ stops growing. The higher the initial energy 
density, the longer the total fireball lifetime until hadronization and 
freeze-out, and the earlier this saturation occurs in the expansion 
history. This makes $v_2$ an {\em early signature} \cite{Sorge} (the 
more so the higher the initial energy density) which is carried by the 
abundant soft final state hadrons.  

\vspace*{-0.5cm}
\hspace*{-1.8cm}
\begin{center}
   \begin{minipage}[t]{7truecm}
         \epsfxsize 7.0truecm \epsfysize 5.6truecm \epsfbox{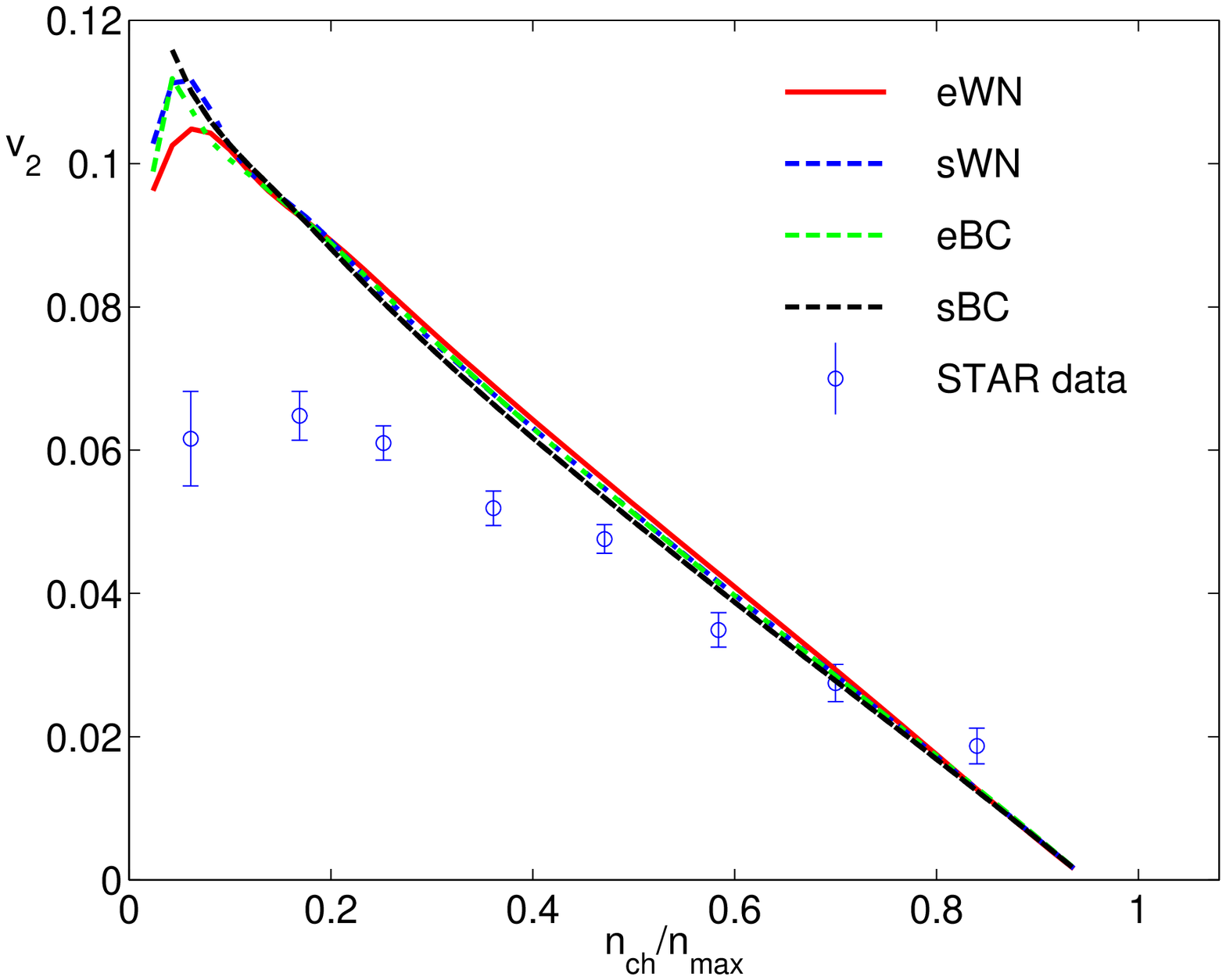}
         \hfill
   \end{minipage}
\vspace*{-0.05truecm}
   \begin{minipage}[t]{7.5truecm}
         \epsfxsize 7.2truecm \epsfysize 5.47truecm \epsfbox{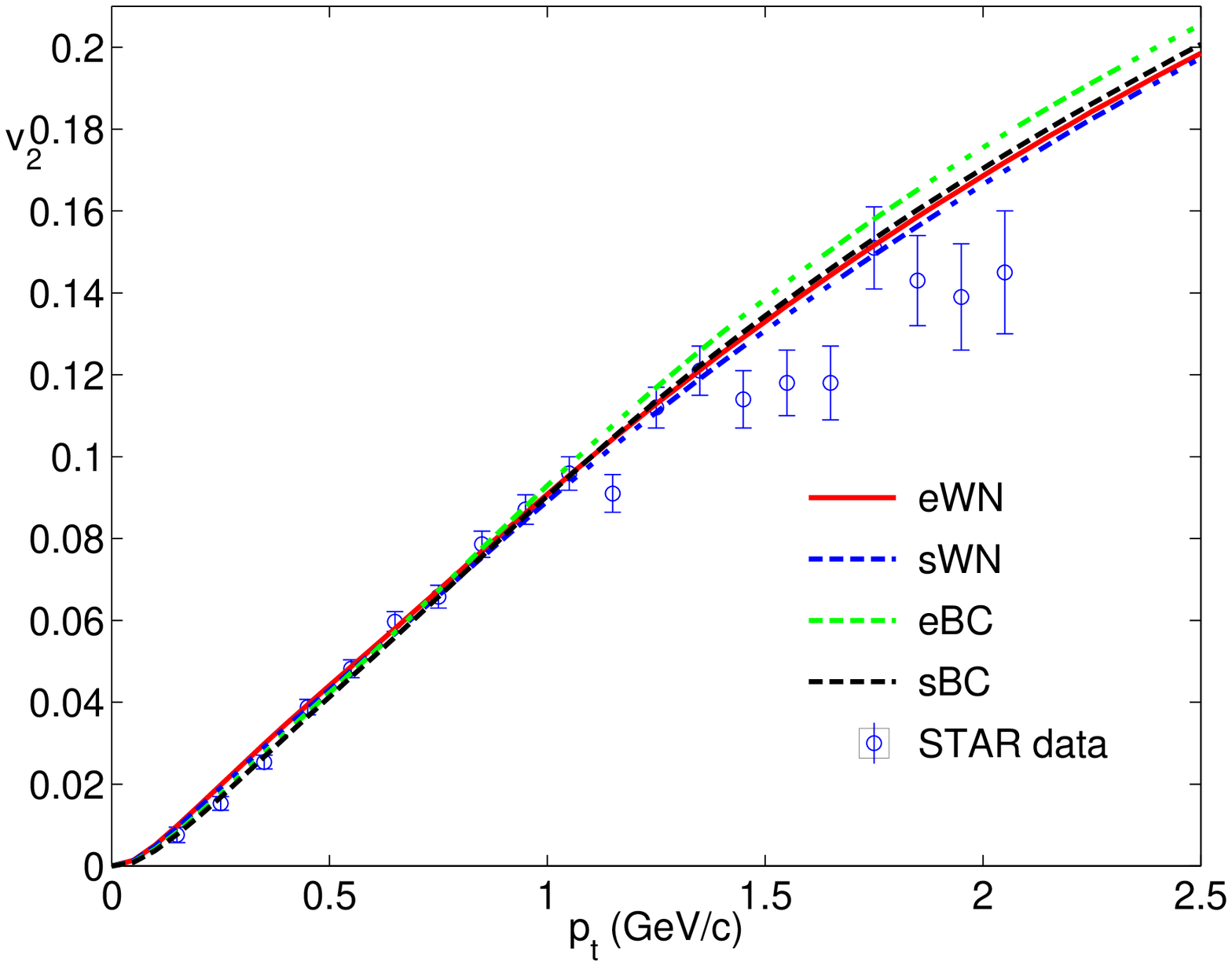}
         \hfill
   \end{minipage}
\end{center}
\vspace*{-0.7cm}
  {\small FIGURE~6. $\pT$ integrated elliptic flow coefficient $v_2$ as 
  function of collision centrality (left) and differential elliptic flow 
  $v_2(\pT)$ for minimum bias collisions (right) for 130\,$A$\,GeV Au+Au 
  collisions at RHIC \protect\cite{STARv2}. The curves are from hydrodynamic 
  collisions with different initial density profiles \protect\cite{KHHET}.}
\label{F6}
\vspace*{0.2cm}

Kinetic simulations \cite{Gyu} show that the momentum-space response to 
the spatial deformation grows monotonically with the interaction strength 
among the constituents of the produced matter. For a fixed spatial 
deformation, the largest $v_2$ response results from a system which is 
coupled so strongly that it thermalizes ``instantaneously'' on fluid
dynamic time scales. This maximal response can thus be computed by solving
the equations of ideal (non-viscous) hydrodynamics \cite{Olli,kolb}.
Fig.~6 shows that the RHIC data exhaust this upper limit for semi-central
Au+Au collisions and low $\pT$, indicating very rapid thermalization on
a time scale of 0.5 fm/$c$ \cite{kolb}. For large impact parameters 
$b{\,>\,}7$\,fm and $\pT{\,>\,}1.5-2$\,GeV/$c$ the measured $v_2$ stays 
below the hydrodynamic prediction, reflecting less efficient thermalization 
in small reaction volumes and at large transverse momenta. The hydrodynamic 
calculations show that the observed agreement between theory and data 
requires thermalization at energy densities well above $\epsilon_{\rm c}$.
It thus becomes difficult to avoid the conclusion that the prehadronic 
state formed in Au+Au collisions at RHIC is indeed a quark-gluon plasma. 

\vspace*{0.2cm}
\begin{center}
{\bf X. \ CONCLUSIONS}
\end{center}
\vspace*{-0.1cm}

Relativistic heavy-ion collisions at the CERN SPS and RHIC have taken 
us into new and unprecedented regions of energy density, almost two 
orders above that of cold nuclear matter and well above the critical 
value for quark deconfinement. We have strong direct experimental 
evidence for a large degree of thermalization and strong collective 
behaviour setting in at a very early collision stage. Elliptic flow
data show that at RHIC most of the anisotropic flow pattern is 
established before the energy density has dropped below $\epsilon_{\rm c}$. 
This requires thermalization in the quark-gluon phase. Even at the SPS,
the observed patterns of chemical equilibrium freeze-out at $T_{\rm c}$
(now confirmed at RHIC), strangeness enhancement and $J/\psi$ suppression 
are fully consistent with predictions based on hypothetical QGP creation.

At the SPS the search for electromagnetic radiation directly emitted from 
the QGP phase did not provide convincing answers. The higher initial 
temperatures and longer plasma lifetimes at RHIC should make that task
a bit easier, but we may still have to wait a while to see results. 
First tentative evidence for jet quenching at RHIC (see \cite{Ullrich})
indicates that with this process we may have found another method of
directly probing the early prehadronic collisions stage. The era of
detailed studies of the properties of the QGP has now begun.

\vspace*{0.5cm}


\end{document}